\documentclass[mypaper,8pt,twoside]{CoAst}
\usepackage{epsf,graphicx,fancyhdr,sfmath}
\renewcommand{\chaptermark}[1]{\markboth{#1}{}}

\newcommand{\kopf}{\small\itshape Comm.\ in Asteroseismology, N$^{\textsf{\underline{o}}}$ 159, 2009\\
Proceedings of the JENAM 2008 Symposium N$^{\textsf{\underline{o}}}$~4:
Asteroseismology and Stellar Evolution}

\newcommand{\Authors}[1]{\begin{center}\normalsize\bf\sf #1 \end{center}}

\renewcommand{\author}[1]{\begin{center}\normalsize\bf\sf #1 \end{center}}
\newcommand{\Address}[1]{\begin{center}\small\sf #1 \end{center}}

\newcommand{\Objects}[1]{{\vspace{3mm}\small \noindent Individual Objects: }\small\sf \hangindent=27truemm \hangafter=1 #1 }

\renewenvironment{abstract}{\section*{Abstract}\normalsize\sf}{}
\newcommand{\References}[1]{\begin{flushleft}{\large References\\}\vspace*{2mm}\small #1 \end{flushleft}}

\newcommand{\chapterCoAst}[2]{\chapter[\sf\normalsize #1\\ \footnotesize \hspace*{5mm}by #2 \sf\normalsize][]{#1\\}\rhead[\fancyplain{}{\sf\footnotesize \center{#1}}]{\fancyplain{}{\sffamily\thepage}}\lhead[\fancyplain{\kopf}{\sffamily\thepage}]{\fancyplain{\kopf}{\sf\footnotesize \center{#2}}}}

\newcommand{\figureCoAst}[5]{\begin{figure}[#4]
\centering
\includegraphics*[#5]{#1}
\caption{#2}
\label{#3}
\end{figure}}

\renewcommand{\chaptername}{}

\newcommand{\acknowledgments}[1]{\vspace*{5mm}\noindent  \textbf{Acknowledgments.} #1}

\def\rfr{\smallskip\par\noindent
        \hangindent=7truemm
        \hangafter=1}

\begin{document}
\sf

\chapterCoAst{TT Arietis -- Observations of a Cataclysmic Variable Star with the MOST Space Telescope}
{J.\,Weingrill, G.\,Kleinschuster, et al.} 
\Authors{J.\,Weingrill$^1$, G.\,Kleinschuster$^2$, R.\,Kuschnig$^3$, J.M.\,Matthews$^4$, A.\,Moffat$^5$, S.\,Rucinski$^6$, D.\,Sasselov$^7$, and W.W.\,Weiss$^3$} 
\Address{
$^1$ Space Research Institute, Austrian Academy of Sciences, Schmiedlstrasse 6, 8042 Graz, Austria,\\
$^2$ Astro Club Auersbach, Wetzelsdorf 33, 8330 Feldbach, Austria,\\
$^3$ Institut f\"ur Astronomie, Universit\"at Wien, T\"urkenschanzstrasse 17, 1180 Wien, Austria,\\  
$^4$ Department of Astronomy \& Physics, UBC, 6224 Agricultural Road, Vancouver, BC V6T 1Z1, Canada,\\
$^5$ Departement de physique, Univ. de Montreal, Montreal, Quebec, QC H3C 3J7, Canada,\\
$^6$ David Dunlap Obs., Dept. of Astronomy, Univ. of Toronto, Toronto, Ontario, L4C 4Y6, Canada,\\
$^7$ Harvard-Smithsonian Center for Astrophysics, Cambridge, Massachusetts, MA 02138, USA
}

\noindent
\begin{abstract}
We measured the photometric flux of the cataclysmic variable TT Arietis (BD+14 341) using the MOST space telescope. Periodic oscillations of the flux reveal the orbital period as well as other features of this binary system. We applied a Discrete Fourier Transform (DFT) on a reduced dataset to retrieve the
frequencies of TT Arietis.
The analysis of the system revealed a photometric period of $3.19$~hours. Though the MOST data has a high cadence of $52.8$~seconds, a fine structure of the accretion disk is not obvious.

\end{abstract}

\Objects{TT Ari}

\section*{Introduction}
The MOST (Microvariability and Oscillations of STars) satellite observed TT Ari between MJD $54395.6$ and $54406.4$. The optical setup of MOST consists of a Maksutov type optical telescope and two identical CCDs. A detailed description of the MOST mission can be found in Walker et\,al.~(2003) and Matthews (2004). The classification of the star is hindered by the low inclination of 20 degrees and therefore mentioned differently in literature. It is most likely to be a VY~Scl-type star {(Wu et\,al.,~2002)} or belongs to the class of SW~Sex stars as mentioned by Kim et\,al.~(2008). The primary photometric period varies  between 3.1824(48) hours and 3.19056(72) hours as listed in Tremko et\,al.~(1996). According to the observations of quasi-periodic oscillations (QPO) between 2005 and 2006, TT~Ari is believed to return from its `positive superhump' state {(Kim et\,al.,~2008)}.
\section*{Methods}
The target was measured with direct imaging photometry because of the low magnitude of the TT Ari system. The initial data reduction has been accomplished by the MOST science team (Rowe, 2006). The first analysis of the data was carried out using discrete Fourier transform (Swan, 1982). The Fourier autocorrelation function (Scargle, 1989) was used as an alternate method, since it shows 36 periods which are not obvious in the original data. In order to remove the remaining signature of the satellite orbit, the first 100~datapoints equivalent to a time lag of 1.452~hours were ignored. The distances between the first three local maxima were averaged to obtain the period from the autocorrelation function. In order to look for QPOs and variations in the periodicity of the signal indicated by the autocorrelation function, a sliding Fourier window analysis (Jacobsen, 2003) was carried out. The Fourier window sizes were scaled from 1.88~hours to 15~hours and a boxcar function was used as a windowing function. The window was shifted over the datapoints without overlapping.
\section*{Results}
The analysis using DFT reveals a period of 3.177(44)~hours with an amplitude of 12.4~mmag. The resulting periodogram can be seen in Figure~\ref{fig:Weingrill_plot}. The main period and the amplitudes of its harmonics correspond to a nearly sinusodial shape of the folded lightcurve. Other methods like the autocorrelation of the data show a slightly different photometric period of 3.194(18)~hours.
\figureCoAst{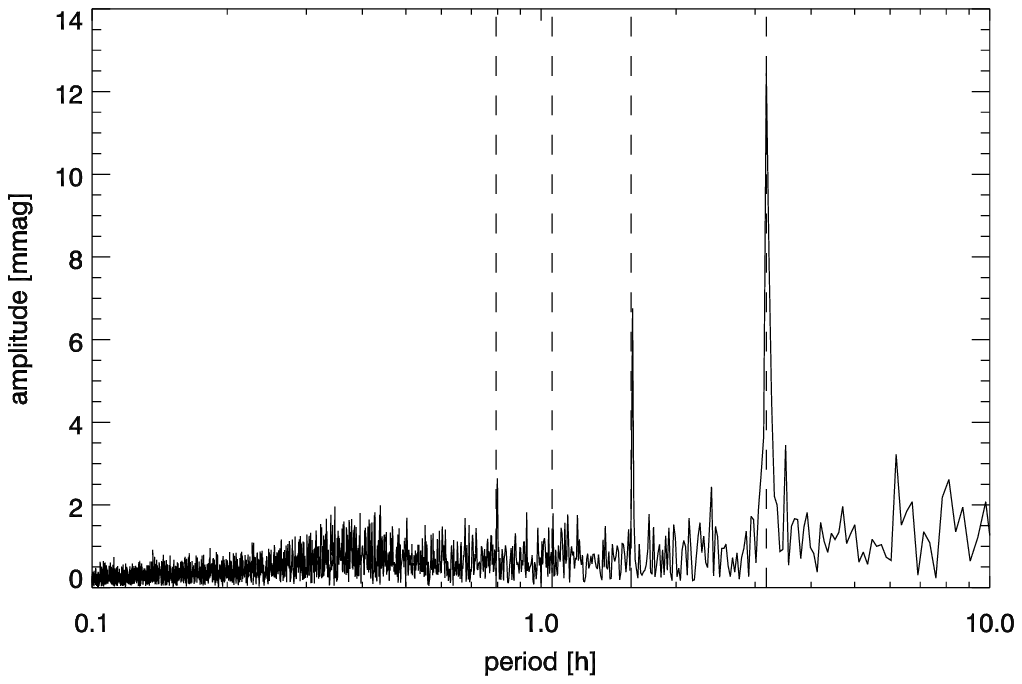}{Periodogram derived from the DFT analysis. Main period and harmonics are indicated by dashed lines.}{fig:Weingrill_plot}{t}{clip,angle=0,width=0.8\textwidth}

The QPOs which can be identified as flickering of the accretion disk is visible in the regime of 19 to 26 minutes. This corresponds to earlier results (see e.g. Semeniuk et\,al., 1987) in a `negative superhump' state. Due to the short observation run of the MOST satellite the current `superhump state' could not be verified.

\acknowledgments{ 
Data was provided by `Universe in a Suitcase -- MOST for all'.
}

\References{
\rfr Walker, G., Matthews, J.M., Kuschnig, R., et\,al.~2003, PASP, 115
\rfr Matthews, J.M., 2004, Bullet. of the AAS, 35, 1563
\rfr Wu, X., Li, Z., Ding, Y., et\,al.~2002, ApJ, 569, 418
\rfr Kim, Y., Andronov, I.L., Cha, S.M., et\,al.~2008, A\&A, accepted
\rfr Tremko, J., Andronov, I.L., Chinarova, L.L. et\,al.~1996, A\&A, 312, 121
\rfr Rowe, J.F. et\,al.~2006, ApJ, 646, 1241-1251
\rfr Swan, R.R. 1982, AJ, 87, 1608
\rfr Scargle, J.D. 1989, ApJ, 343, 874
\rfr Jacobsen, E. \& Lyons, R. 2003, SP-M, 74
\rfr Semeniuk I., Schwarzenberg-Czerny A., Duerbeck H. et~al. 1987, A\&A, 37, 197
}

\end{document}